\documentclass[aps,prb,reprint,superscriptaddress]{revtex4-2}

\usepackage{graphicx}
\usepackage{color}
\usepackage{dcolumn}  
\usepackage{braket}
\usepackage{hyperref}
\usepackage{amsmath}
\usepackage{siunitx}
\usepackage{chemformula}
\begin{document}

\def\tc{\ensuremath{T_{\rm C}}}
\def\lsco{\ch{La_{0.3}Sr_{0.7}CoO_{3-$\delta$}}}



\title{
	\boldmath
Stabilization of the oxygen concentration in \lsco\ thin films by \ch{LaAlO3} capping layer \unboldmath}

\author{M. Kiaba}
\email[]{kiaba@mail.muni.cz}
\affiliation{Department of Condensed Matter Physics, Faculty of Science, Masaryk University, Kotl\'a\v{r}sk\'a 2, 611 37 Brno, Czech Republic}

\author{O. Caha}
\affiliation{Department of Condensed Matter Physics, Faculty of Science, Masaryk University, Kotl\'a\v{r}sk\'a 2, 611 37 Brno, Czech Republic}

\author{F. Abadizaman}
\affiliation{Department of Condensed Matter Physics, Faculty of Science, Masaryk University, Kotl\'a\v{r}sk\'a 2, 611 37 Brno, Czech Republic}

\author{A. Dubroka}

\affiliation{Department of Condensed Matter Physics, Faculty of Science, Masaryk University, Kotl\'a\v{r}sk\'a 2, 611 37 Brno, Czech Republic}


\date{\today}

\begin{abstract}
We have grown \ch{La_{0.3}Sr_{0.7}CoO_{3-$\delta$}} thin films by pulsed laser deposition on \ch{(LaAlO3)_{0.3}(Sr2TaAlO3)_{0.7}} substrates with and without a protective LaAlO$_3$ capping layer and investigated their structural and magnetic properties. We have observed that, in the uncapped films, the Curie temperature strongly decreased after annealing in  helium atmosphere, and it significantly decreased even in samples stored for several weeks at room temperature. The decrease of the Currie temperature is caused by an increase of the concentration of oxygen vacancies, $\delta$. However, we show that already a 3~nm  thin \ch{LaAlO3} capping layer can essentially conserve $\delta$\ at room temperature, and it considerably slows down the formation of oxygen vacancies at elevated temperatures. 
\end{abstract}
\pacs{xxx}
\maketitle

\section{Introduction}
Transition metal oxides with the perovskite structure have a wide spectrum of properties such as superconductivity, magnetism or ferroelectricity and have potential applications in future electronics \cite{Coll2019}. One way to tune these properties is to change the concentration of oxygen vacancies that can, e.g., turn a semiconductor into a superconductor \cite{Barbut2002}, a metal into an insulator \cite{Lebon2004} or an antiferromagnetic order into a ferromagnetic one~\cite{Toepfer1997}. The control of the concentration of oxygen vacancies is thus an important aspect of a preparation of transition metal oxides. 

\begin{figure}[t]
	\centering
	\vspace*{0cm}
	\hspace*{-0.5cm}
	\includegraphics[width=7cm]{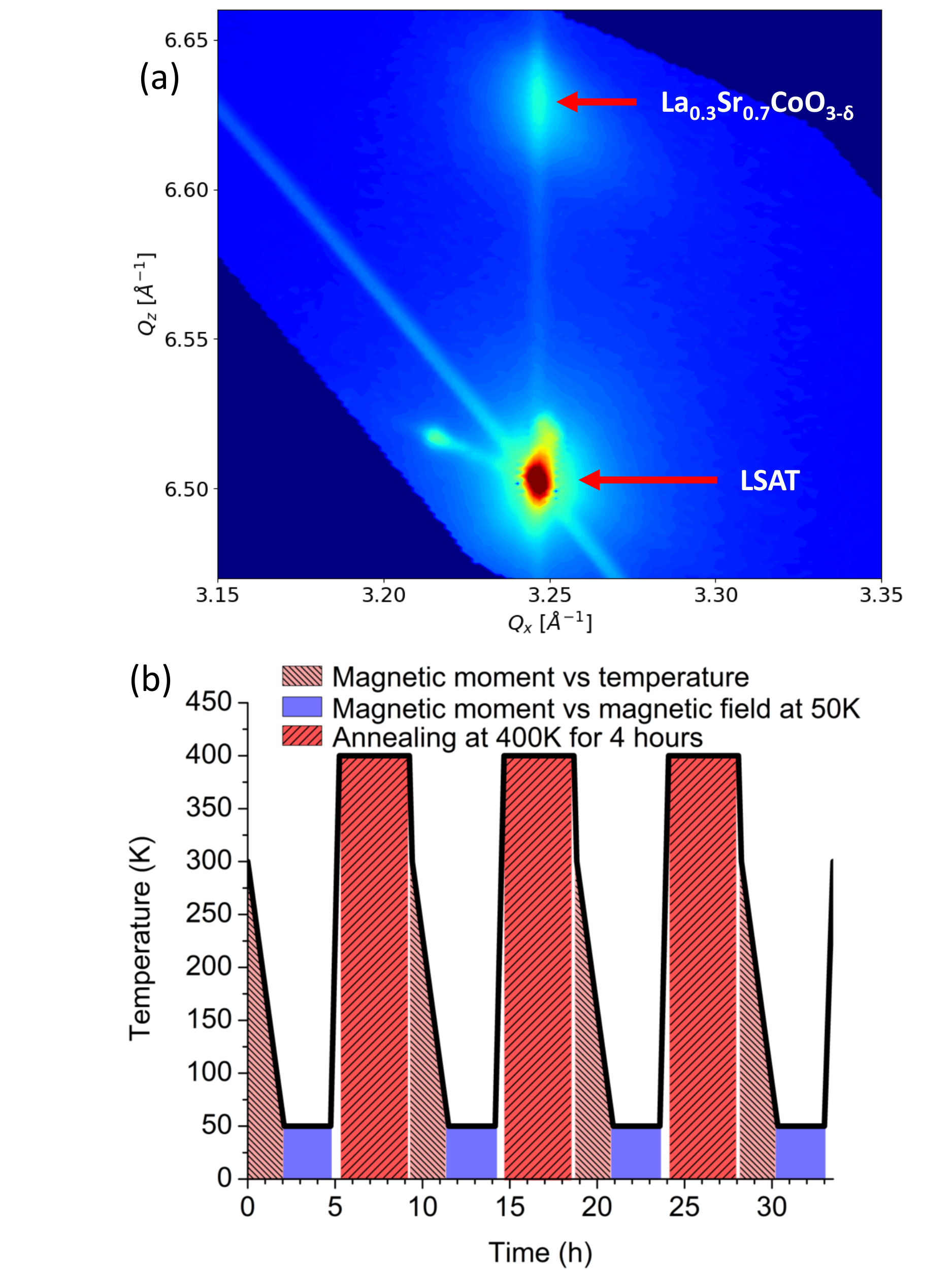}
	\vspace*{-0.2cm}
	\caption{(a) Reciprocal space map of \lsco\ film grown on LSAT substrate measured in the vicinity of 204 reciprocal lattice point. The arrows denote the diffraction peaks from the substrate and the film, respectively. (b) The time scheme of the measurement-annealing cycles performed in \SI{20}{\milli\bar} He atmosphere. }
	\label{maps}
\end{figure}

Compounds such as La$_{1-x}$Sr$_{x}$CoO$_3$ and La$_{1-x}$Sr$_{x}$FeO$_3$ exhibit an instability of oxygen content particularly for high Sr fraction, i.e.,
for $x \geq 0.5$~\cite{Enriquez2016,Perret2017,Chennabasappa2020}. We have observed a significant decrease of the Curie temperature (\tc) of \lsco\ (LSCO) films stored at room conditions for several weeks due to the formation of oxygen vacancies. To prevent or at least slow down this process, we have capped the LSCO films with \ch{LaAlO3} capping layers with various thicknesses. We have chosen \ch{LaAlO3} because of its chemical stability and the large band-gap, which potentially allows the investigation of optical properties in the visible and ultraviolet spectral range. It turned out that only a 3~nm thin capping layer essentially stabilizes the oxygen content at room temperature and significantly slows down the creation of oxygen vacancies at higher temperatures. Using a capping layer as prevention of a change in oxygen content was previously reported for \ch{EuTiO3} films~\cite{Shimamoto2013} and SrFeO$_{3-\delta}$ films~\cite{Enriquez2016}.

\section{Experiment}
\label{Experiment}

\begin{figure*}[t]
	\centering
	\vspace*{-0.1cm}
	\hspace*{-0.5cm}
	\includegraphics[width=17cm]{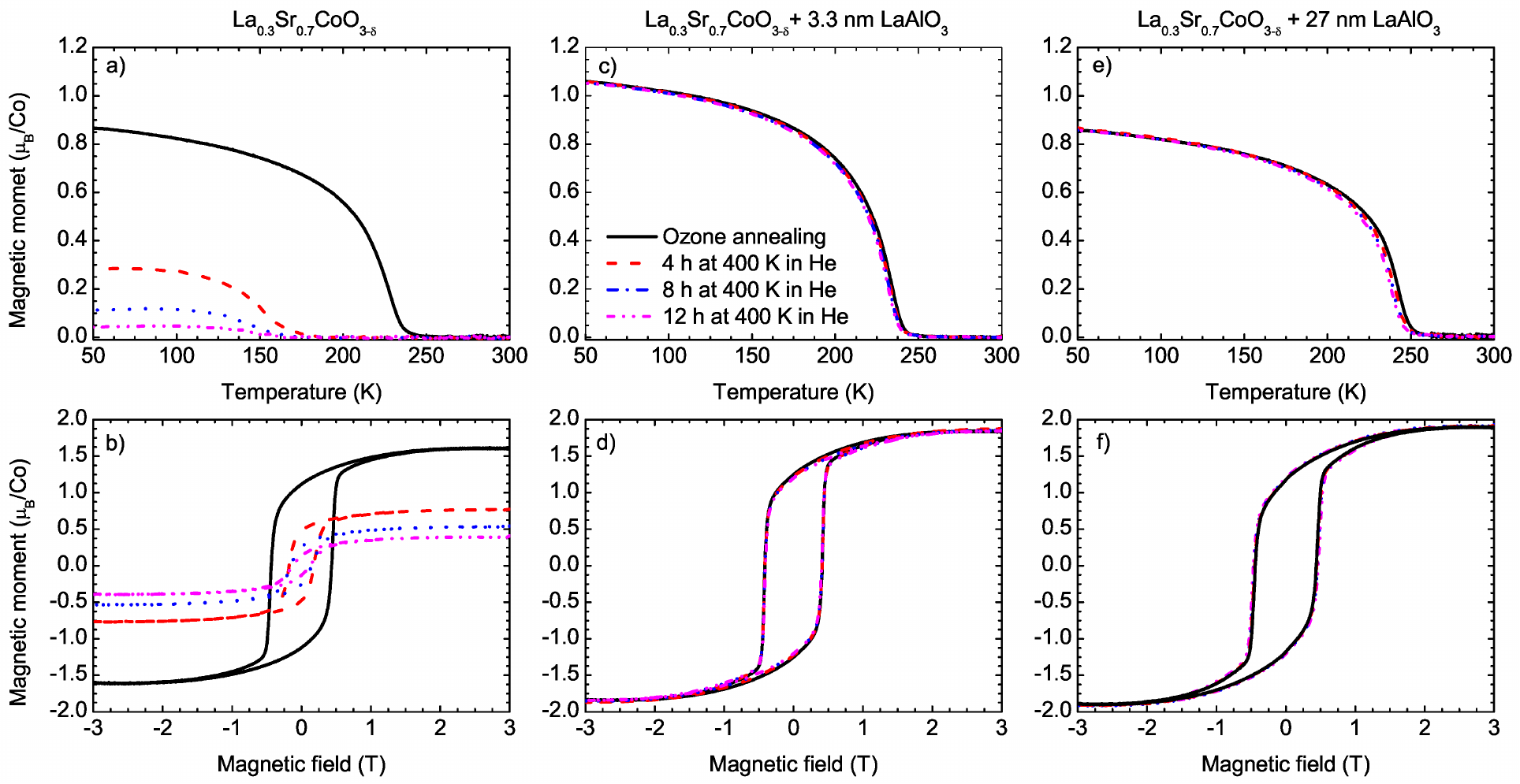}
	\vspace*{-0.2cm}
	\caption{ Magnetic moment of \lsco\ film (left panels), \lsco\ film capped with  \SI{3}{\nano\meter} thin \ch{LaAlO3} layer (center panels), and \lsco\ film capped with 27~nm thin \ch{LaAlO3} layer (right panels). The measurements of the magnetic moment were performed as a function of temperature at \SI{10}{m\tesla} (upper panels) or as a function of the applied magnetic field at \SI{50}{\kelvin} (lower panels). The lines correspond to measurements after various annealing steps as described in the legend.}
	\label{magprop}
\end{figure*}

LSCO thin films with and without \ch{LaAlO3} capping layer were fabricated by the pulsed laser deposition technique. KrF laser with a wavelength of \SI{248}{\nano\meter} was focused on a target with a fluency of \SI{1.8}{\joule\per\square\centi\meter} over an area of about \SI{1}{\square\milli\meter}. The distance of the target from the substrate was set to \SI{55}{\milli\meter}. LSCO films were epitaxially grown on 10~$\times$~10~mm$^2$ \ch{(LaAlO3)_{0.3}(Sr2TaAlO3)_{0.7}} (LSAT) substrates heated with an infrared laser to \SI{700}{\celsius} and the temperature was measured with an infrared pyrometer. The 
growth was performed in an oxygen atmosphere with a partial pressure of \SI{0.1}{\milli\bar}, and using pulse frequency of \SI{7}{\hertz}. For the deposition of the \ch{LaAlO3} capping layer, the substrate temperature was increased to \SI{800}{\celsius}, and pulse frequency was set to  \SI{1}{\hertz}.
The thickness of the LSCO films of $(27\pm1)$\,nm was determined by x-ray reflectivity. We have studied three samples: one without any capping, and two samples with  \SI{3}{\nano\meter} and \SI{27}{\nano\meter} thin \ch{LaAlO3} capping layers, respectively. To reduce the concentration of oxygen vacancies, the samples were annealed {\it ex-situ} for \SI{3}{\hour} at \SI{400}{\celsius} in an ozone atmosphere. The annealing in ozone is an important step that reduces the concentration of oxygen vacancies significantly more than  annealing in pure oxygen. Structural characterization of the samples was performed by x-ray diffraction. Magnetic properties of the samples were measured using the vibrating sample magnetometry (VSM) (VersaLab, Quantum Design).

\section{Data analysis and discussion}
Figure \ref{maps}(a) shows the reciprocal space map measured in the vicinity of 204 diffraction of the uncapped LSCO film. The arrows denote the diffraction peaks from the LSAT substrate, and the LSCO film, respectively. The diffraction peak from the LSCO film is located at the same $Q_x$ position as the one from the substrate, demonstrating that the film is pseudomorphic. LSAT has a cubic lattice with a lattice parameter of \SI{3.868}{\angstrom}. Bulk LSCO has the cubic structure~\cite{Wu2003}, however, LSCO films grown on LSAT substrate are under a tensile strain and consequently they exhibit a small tetragonal distortion. 

\begin{figure*}[t]
	\centering
	\vspace*{-0.1cm}
	\hspace*{0cm}
	\includegraphics[width=12cm]{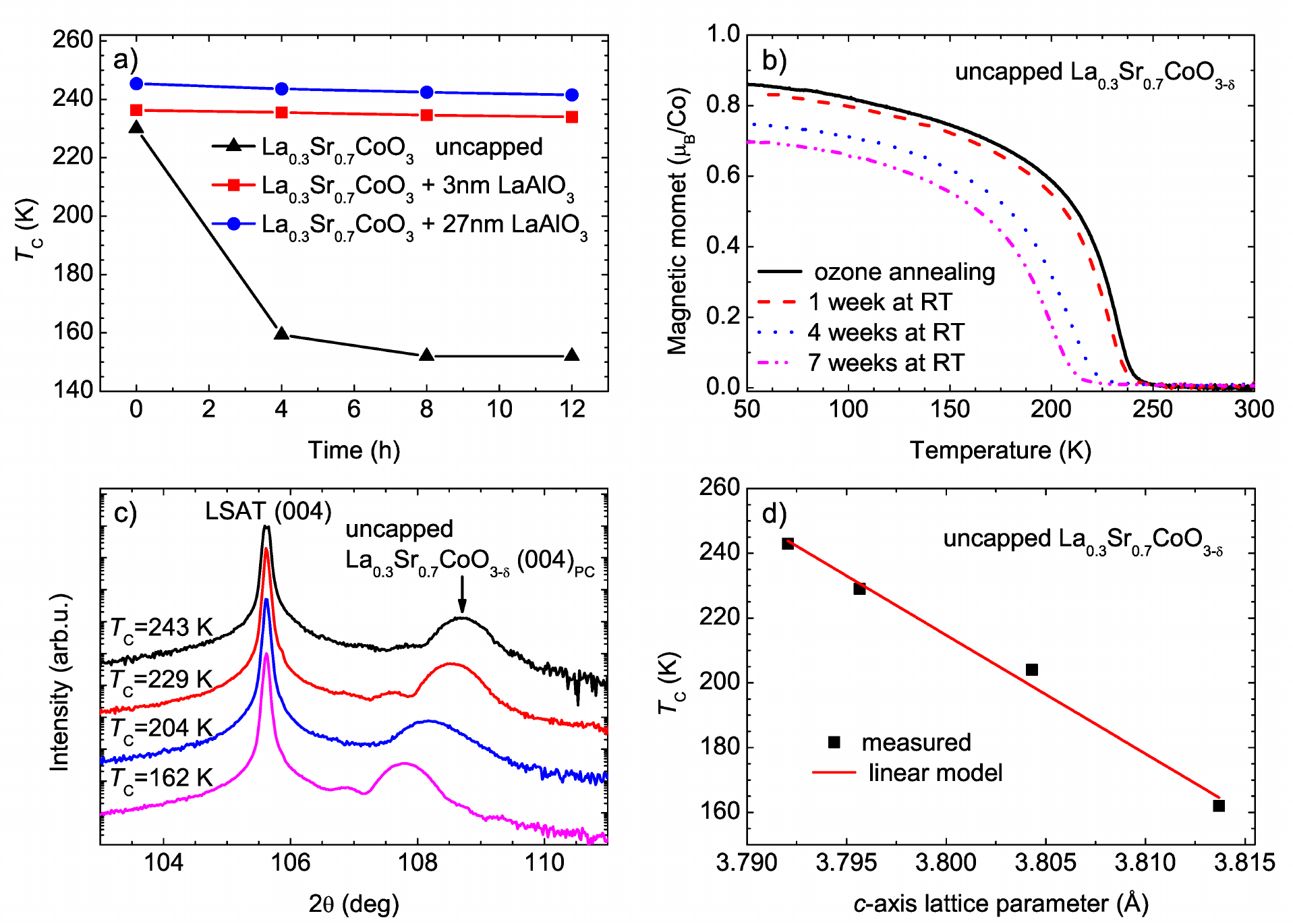}
	\vspace*{-0.2cm}
	\caption{a) Values of \tc\ of \lsco\ films with and without \ch{LaAlO3} capping layer (see the legend) as a function of annealing time at \SI{400}{\kelvin} in He.  b) Magnetic moment with respect to temperature of the uncapped \lsco\ film measured at \SI{10}{m\tesla} at different times after the ozone annealing while storing the sample in air at room temperature (RT). c) X-ray diffraction near 004 pseudocubic (pc) lattice point of the  uncapped \lsco\ film with various oxygen content and corresponding values of \tc\ as denoted in the figure. The spectra are vertically shifted for clarity. d) Curie temperature versus $c$-axis lattice parameter determined from the data of panel c). The line represents the fit with a linear function. }
	\label{decrease}
\end{figure*}

To compare the stability of the samples, we have performed an oxygen-reducing annealing experiment in helium atmosphere at a pressure of about \SI{100}{\milli\bar}, which is the standard measurement condition in the VSM Versalab chamber. The time scheme of the experiment is shown in Fig.~\ref{maps}(b). In a short time after the annealing in ozone, the temperature dependence of the magnetic moment was measured from \SI{300}{\kelvin} to \SI{50}{\kelvin} in a magnetic field of 10\,mT followed by the measurement of the hysteresis loop at \SI{50}{\kelvin} in magnetic fields up to $\pm$ \SI{3}{\tesla}. Consequently, the sample was heated to \SI{400}{\kelvin}, where it was held for \SI{4}{\hour}, and cooled to \SI{300}{\kelvin} with a ramp rate of \SI{12}{\kelvin/\minute}. After this He annealing, we again performed the same measurements of the temperature dependence of the magnetic moment between \SI{300}{\kelvin} and \SI{50}{\kelvin} and of the hysteresis loop at \SI{50}{\kelvin}. This cycle was repeated two more times. 

The magnetic moment of all samples, expressed in terms of the number of Bohr magnetons per cobalt ion, are shown in Fig. \ref{magprop}.  Figure~\ref{magprop}(a) demonstrates that \tc\ of the film without the capping layer decreased substantially from about \SI{230 }{\kelvin} to about \SI{160}{\kelvin} after the first annealing cycle, and it continued to decrease after the subsequent cycles. 
Figure~\ref{magprop}(b) shows the hysteresis loops measured at 50\,K. The diamagnetic contribution of the substrate was subtracted from the data using a linear fit of the magnetic moment in the high-field region. The hysteresis loops demonstrate that the strong reduction in \tc\ is accompanied by a strong reduction of the saturated magnetic moment. Since it is possible to restore the \tc\ of the sample to the original value by the ozone annealing (data not shown), it is clear that this strong reduction of \tc\ and of the saturated magnetic moment is due to the oxygen loss occurring through the surface. 
Similarly, a loss of oxygen at elevated temperatures was reported for \ch{La_{0.7}Sr_{0.3}MnO3} thin films~\cite{Cao2018}. The results of the magnetic moment measurements with respect to temperature and magnetic field for the LSCO film with \SI{3}{\nano\meter} thin \ch{LaAlO3} capping layer are shown in Fig.~\ref{magprop}(c) and \ref{magprop}(d), respectively. Notably, the strong reduction of \tc\ and of the saturated magnetic moment observed in the uncapped film is essentially absent, with only comparably a very weak reduction in \tc\  after the He annealing. 
The results demonstrate that already a \SI{3}{\nano\meter} thin \ch{LaAlO3} capping layer substantially slows down  the oxygen loss. 
Figures~\ref{magprop}(e) and \ref{magprop}(f) show the data for the LSCO film with \SI{27}{\nano\meter} thin \ch{LaAlO3} capping layer also exhibiting the relative stability with respect to annealing and reaching even higher \tc\ than the other samples. 
Analogically to our results, a stabilization of oxygen concentration in thin films of SrFeO$_{3-\delta}$, that also tend to be oxygen deficient, was achieved with a 300~nm thick amorphous Al$_2$O$_3$ passivation layer~\cite{Enriquez2016}. Our results demonstrate that the stabilization of LSCO films in particular, and potentially films of perovskite oxides in general,  is possible with a much thinner epitaxial \ch{LaAlO3} layer that allows, e.g.,  optical investigations and/or subsequent epitaxial growth.

To quantitatively compare the samples, we have determined \tc\  using a linear fit to the square of the magnetic moment  with respect to temperature just below \tc\ (fits not shown). The obtained values of \tc\ of all samples are summarized in Fig.~\ref{decrease}(a). The values of \tc\ of the capped samples decreased by only about  \SI{2}{\kelvin} after all three He annealing cycles. The sample with 27~nm \ch{LaAlO3} capping layer yielded the highest \tc\ among our samples of  \SI{245}{\kelvin}. Even after three He annealing cycles, its \tc\ is higher than in the other samples. 
The value of $\tc=245$\,K is slightly higher than values reported by \cite{Wu2003,Sunstrom1998} in bulk LSCO but considerably lower than \SI{280}{\kelvin} recently achieved by M. Chennabasapp {\it et al.} in bulk samples treated by the electrochemical oxidation~\cite{Chennabasappa2020}. We are unaware of any reported \tc\ in LSCO thin films. 
The reason why the highest \tc\ of our samples is still considerably lower than the record value of \SI{280}{\kelvin} is either due to a residual oxygen vacancy concentration $\delta$  or/and due to the tensile strain imposed by the substrate. The maximum saturated moment of our films at \SI{50}{\kelvin}, see Figs.~\ref{magprop}(b), \ref{magprop}(d) and \ref{magprop}(f), increases with increasing capping layer thickness, i.e., from $1.60 \pm 0.03 \mu_B$/Co for bare film to $1.84 \pm 0.03 \mu_B$/Co for \SI{3}{\nano\meter} thin capping layer and to $1.90 \pm 0.06 \mu_B$/Co for film with \SI{27}{\nano\meter} thin capping layer. The values of the capped films agree within the error-bars with the value of $1.85 \mu_B$/Co achieved at \SI{5}{\kelvin} in bulk samples treated by the electrochemical oxidation~\cite{Chennabasappa2020}.

We have investigated in more detail the instability of oxygen content in the uncapped sample stored in air at room temperature over a longer period. Magnetic moment as a function of temperature measured at various time after the annealing in ozone is shown in Fig.~\ref{decrease}(b). The value of \tc\ decreased by 4, 26 and  \SI{31}{\kelvin} in 1, 4 and 7 weeks, respectively, after the annealing, demonstrating that the uncapped LSCO film significantly looses oxygen even at room conditions. The saturated magnetic moment decreased similarly to the annealing experiment shown in Fig.~\ref{magprop}(a). Remarkably, we did not observe any detectable change of \tc\ in the capped films stored for one month at room temperature. An unstable behavior at room temperature with a smaller change of \tc\ was observed in bulk \ch{La_{0.2}Sr_{0.8}CoO3} after 8 months~\cite{Chennabasappa2020}. Obviously, the loss of oxygen occurs much faster in films than in bulk due to a much higher surface-to-volume ratio.

The loss of oxygen content can also be observed by the change of the lattice parameter~\cite{Enriquez2016,Cao2018}. Figure~\ref{decrease}(c)  shows several x-ray diffraction spectra of the uncapped LSCO films with various oxygen content measured near the 004 diffraction. One spectrum was acquired right after the annealing in ozone for \SI{5}{\hour} (\tc\ of \SI{243}{\kelvin}), two of them after various times at room conditions after the annealing in ozone (\tc\ of 229 and \SI{204}{\kelvin}) and one of them after the He annealing (\tc\ of  \SI{162}{\kelvin}). 
The diffraction spectra exhibit a strong 004 diffraction peak from the LSAT substrate at about \ang{105.6}, and 004 diffraction from the LSCO film with the angular position increasing from about \ang{107.8} to  about \ang{108.8} with increasing \tc. Figure~\ref{decrease}(d) shows the dependence of \tc\ versus the $c$-axis lattice parameter, $a_c$,  obtained from the angular position of the 004 diffraction. The dependence is essentially linear and the fit of a linear function $\tc=A+B a_c$ to the data yielded $A=(14130\pm1000)$\,K and $B=(-3660\pm250)$\,K/\AA.

\section{Conclusion}
\label{Conclusion}

We have prepared epitaxial \lsco\ thin films on LSAT substrates with and without \ch{LaAlO3} capping layer. The samples were annealed in ozone to reduce the concentration of oxygen vacancies. 
We observed that the value of \tc\ of the uncapped film was strongly reduced due to the loss of oxygen after the annealing in  He atmosphere at \SI{400}{\kelvin}. The film was unstable even at room conditions where \tc\ decreased by \SI{31}{\kelvin} after 7 weeks from the annealing in ozone. Notably, only  \SI{3}{\nano\meter} thin \ch{LaAlO3} capping layer essentially stabilizes the film at room conditions and significantly slows down the oxygen loss in  He atmosphere at \SI{400}{\kelvin}. The highest \tc\ of \SI{245}{\kelvin} was achieved in the film with \SI{27}{\nano\meter} \ch{LaAlO3} capping layer which is, to our best knowledge, the highest \tc\ observed in  a \lsco\ thin films.

\begin{acknowledgments}
This work was financially supported by the MEYS of the Czech Republic under the project  CEITEC 2020 (LQ1601), by the Czech Science Foundation (GA\v{C}R) under Project No. GA20-10377S, and by the Operational Programme Research, Development and Education - Project ``Postdoc2MUNI" (No. CZ.02.2.69/0.0/0.0/18\_053/0016952). CzechNanoLab project LM2018110 funded by MEYS CR is gratefully acknowledged for the financial support of the measurements/sample fabrication at CEITEC Nano Research Infrastructure. 
\end{acknowledgments}

\bibliographystyle{apsrev4-2}
\bibliography{Cobaltates}
\end{document}